\documentclass[12pt]{elsarticle}
\usepackage{latexsym, amsmath,setspace,fullpage, titlesec}
\usepackage{natbib}
\usepackage[multiple]{footmisc}
\setcitestyle{authoryear, aysep={}}
\setcitestyle{round}
\setcitestyle{semicolon}
 
\usepackage[title]{appendix}
\usepackage{chngcntr}
\usepackage[english]{babel}
\usepackage{rotating}

\usepackage{threeparttable}
\usepackage{sectsty}
\usepackage{mathtools}

\sectionfont{\normalsize\bfseries}
\subsectionfont{\normalsize\normalfont}
\subsubsectionfont{\normalsize\normalfont\em}
\usepackage{caption}
\begin{document}
\hspace{.2 in}
\begin{flushleft}
\thispagestyle{empty}
\textbf{Unified representation of the C3, C4, and CAM photosynthetic pathways with the Photo3 model}

\vspace{.25cm}
Samantha Hartzell,$^{1}$$^{,}$$^{2}$ Mark S. Bartlett,$^{3}$  and Amilcare Porporato$^{1}$$^{,}$$^{2\ast}$
\end{flushleft}

\vspace{.3cm}
\textbf{This is a preprint.} The final, published version may be found in Ecological Modelling at https://doi.org/10.1016/j.ecolmodel.2018.06.012 and may be cited as:

\vspace{.2cm}
Hartzell S, Bartlett, M, Porporato A (2018) "Unified representation of the C3, C4, and CAM photosynthetic pathways with the Photo3 model" Ecological Modelling (384):173-187. DOI: 10.1016/j.ecolmodel.2018.06.012
\vspace{1.5cm}
\begin{center}
$^\ast$Corresponding Author: Amilcare Porporato\\
Department of Civil and Environmental Engineering\\
Princeton Environmental Institute\\
Princeton University\\
E-208 E-Quad, Princeton, NJ 08540 U.S.A.\\
Phone +1 609 258 2287\\
e-mail aporpora@princeton.edu
\end{center}
\vspace{1cm}
\normalsize{$^{1}$Department of Civil and Environmental Engineering, Princeton University}\\
\normalsize{$^{2}$Princeton Environmental Institute, Princeton University}\\
\normalsize{$^{3}$Department of Civil and Environmental Engineering, Duke University}\\
\newpage
\pagenumbering{arabic}
\pagestyle{plain}


\section*{Abstract}

Recent interest in crassulacean acid metabolism (CAM) photosynthesis has resulted in new, physiologically based CAM models. These models show promise, yet unlike the more widely used physiological models of C3 and C4 photosynthesis, their complexity has thus far inhibited their adoption in the general community. Indeed, most efforts to assess the potential of CAM still rely on empirically based environmental productivity indices, which makes uniform comparisons between CAM and non-CAM species difficult. In order to represent C3, C4, and CAM photosynthesis in a consistent, physiologically based manner, we introduce the Photo3 model. Photo3 unites a common photosynthetic and hydraulic core with components depicting the circadian rhythm of CAM photosynthesis and the  carbon-concentrating mechanism of C4 photosynthesis. This work allows consistent comparisons of the three photosynthetic types for the first time. It also allows the representation of intermediate C3-CAM behavior through the adjustment of a single model parameter. Model simulations of \textit{Opuntia ficus-indica} (CAM), \textit{Sorghum bicolor} (C4), and \textit{Triticum aestivum} (C3) capture the diurnal behavior of each species as well as the cumulative effects of long-term water limitation. These results show the model's potential for evaluating the tradeoffs between C3, C4, and CAM photosynthesis, and for better understanding CAM productivity, ecology, and climate feedbacks.

\section*{Keywords}

C3 photosynthesis $\cdot$ C4 photosynthesis $\cdot$ Crassulacean acid metabolism (CAM) $\cdot$  Plant water storage $\cdot$  Soil-plant-atmosphere continuum

\section{Introduction}

Crassulacean acid metabolism (CAM) and C4 photosynthesis are thought to have evolved as add-ons to the classical C3 photosynthetic pathway around 20-30 million years ago \citep{Keeley2003}. Both photosynthetic processes achieve increased water use efficiency by concentrating CO$_2$ at the site of the dark reactions of photosynthesis. Today, C4 and CAM plants fill important ecological niches in grasslands, rainforests, and arid ecosystems; C4 plants dominate in grasslands where they account for almost 25\% of terrestrial primary production \citep{Still2003} whereas CAM plants make up almost 50\% of plant biomass in certain arid and semi-arid regions of the world \citep{Syvertsen1976}. C4 crops such as corn (\textit{Zea mays}), sugarcane (\textit{Saccharum officinarum}), and sorghum (\textit{Sorghum bicolor}) comprise 22\% of the eighteen most common crops \citep{Leff2004}. CAM crops such as prickly pear (\textit{Opuntia ficus indica}), agave (\textit{Agave tequilana}), and pineapple (\textit{Ananas comosus}) are also economically significant, particularly in arid and semi-arid regions of the world. Due to their extremely high water use efficiency and heat tolerance, the potential of CAM plants for food, fodder, and biofuel production will only become more significant as climate uncertainty and tensions over food scarcity increase \citep{Nobel1991, GarciadeCortazar1992, Borland2009, Owen2014, Mason2015}.

Despite the prevalence and importance of CAM plants, physiological modeling of CAM photosynthesis is well behind that of C3 and C4 photosynthesis \citep{Farquhar1980, VonCaemmerer1999}. Indeed, physiological models of CAM have only recently been introduced (see e.g. \citet{Owen2013, Bartlett2014}) and have not been widely adopted. Instead, Nobel's Environmental Productivity Index (EPI), introduced in 1984, is the standard method of predicting net carbon uptake and yield for CAM plants \citep{Nobel1988, Nair2012}. This index, based on multiplicative indices for water, temperature and photosynthetically active radiation (PAR), is entirely empirical and does not include a physiological representation of the CAM process. Furthermore, the index is designed to be calculated at a timescale of one month \citep{Nobel1988, Owen2014}, thus failing to take into account environmental variability at daily and weekly timescales, which has been shown to be an important factor in CAM functioning \citep{Bartlett2014, Hartzell2015}. This hampers the assessment of the potential impacts of CAM plants on climate, agriculture, and bioenergy production. Most climate modeling efforts include land surface models with a physiologically based representation of C3 \citep{Rogers2017}, and, often C4 photosynthesis \citep{Cox2001, Cowling2007, Milly2014}, but none currently include CAM photosynthesis, an important component of dryland and tropical ecosystems. The recent push for physiologically based crop modeling has also failed to take CAM crops into account. Multiple existing crop models, including 2Dleaf and MCWLA, are based on physiological models of C3 photosynthesis and stomatal conductance \citep{Pachepsky1996, Tao2009}, and the GECROS model supports both C3 and C4 photosynthesis based on modifications to the Farqhuar model \citep{Yin2005}. Despite these advances, no crop models currently exist that are capable of coherently representing the three photosynthetic types. This discrepancy propagates into the analysis of bioenergy potential. While detailed biophysical models of C3 and C4 crops enable analyses of their potential for bioenergy production \citep{Miguez2012, Nair2012}, lack of detailed CAM modeling poses a problem in better understanding its potential in this area \citep{Yan2011, Nair2012, Owen2014, Davis2015}. 

The Photo3 model addresses this need by providing a consistent, physiologically based description of CAM, C3, and C4 photosynthesis coupled to environmental conditions. The model seeks to balance the complexity required to faithfully represent each process with simplicity and clarity. To achieve this, the model leverages the commonalities between the three photosynthetic types. The core of the model is based on the Rubisco-mediated carbon assimilation achieved by the light reactions and Calvin cycle of C3 photosynthesis. When representing CAM and C4 plants, this core is combined with a model for carbon fixation via phosphoenolpyruvate carboxylase (PEPC). The method for representing CAM plants is based on \citet{Bartlett2014}, and adds a component for malic acid storage and release which is governed by an endogenous circadian rhythm. When representing C4 plants, a carbon concentrating mechanism based on \citet{Collatz1992, VonCaemmerer1999, Vico2008} is added to the model core. The resulting integrated model allows plants of all three photosynthetic types to be simulated on a consistent basis and in a wide variety of soil and atmospheric conditions. 

In this work, the model is parameterized for one representative species from each photosynthetic type: \textit{Opuntia ficus-indica} (CAM), \textit{Triticum aestivum} (C3), and \textit{Sorghum bicolor} (C4). Stomatal conductance, carbon assimilation, and water use of the three species are compared at the daily and monthly scale. Finally, intermediate C3-CAM behavior is explored through the adjustment of CAM model parameters. The Photo3 model accurately captures a wide range of photosynthetic behaviors and shows promise for applications in ecological, climate, and crop modeling. Written in Python, the model is open source and publicly available on GitHub. It employs a modular structure which allows it to be easily integrated with other routines for use in a variety of applications.

\section{Materials and Methods}

\subsection{Overview of the Photo3 model}

\begin{figure}
	\centering
	\includegraphics[width=15 cm]{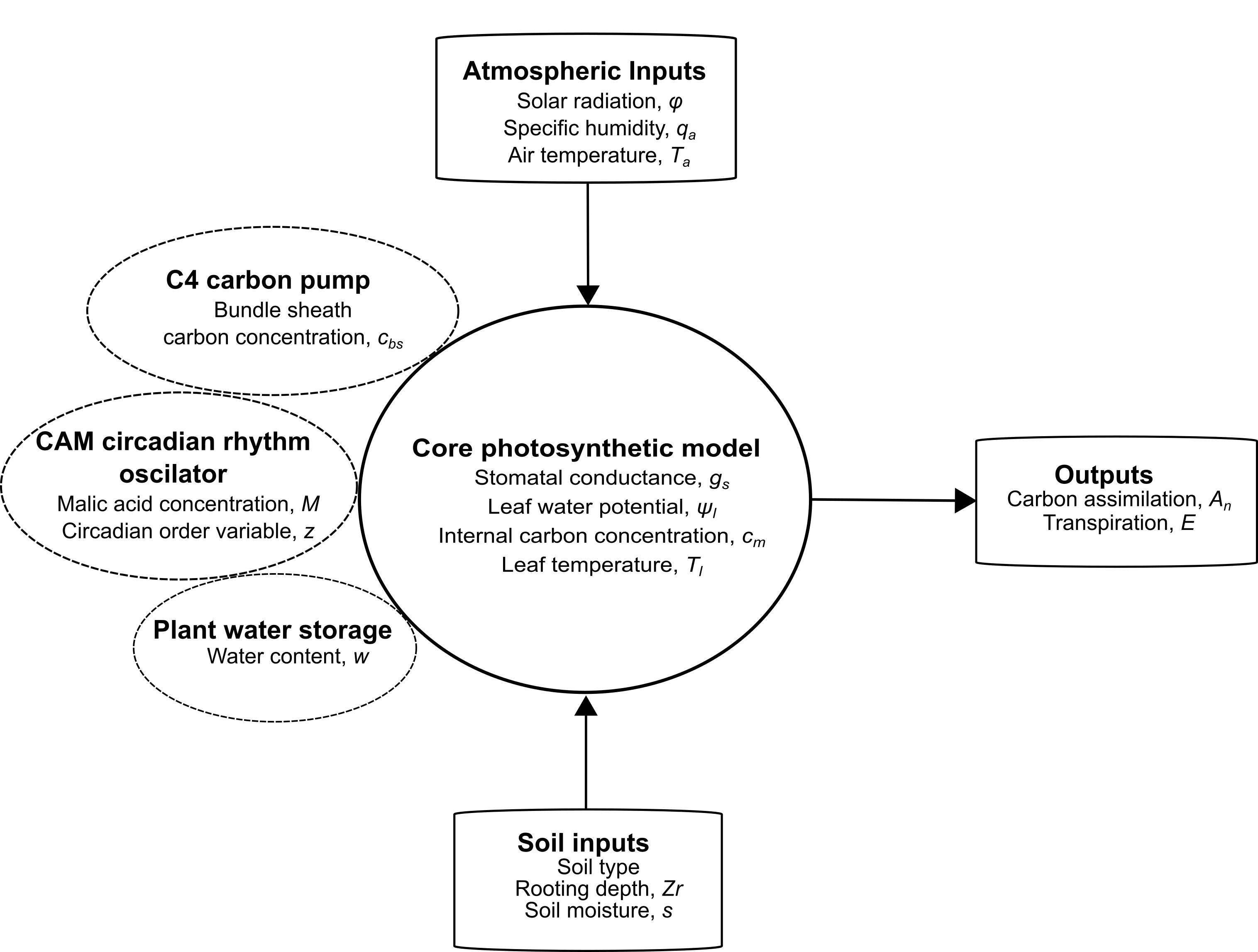}
	\caption{{\bf Photo3 model schematic.} The Photo3 model is based on a core model of C3 photosynthesis with options to represent C4 photosynthesis, CAM photosynthesis, and plant water storage.}
	\label{fig:modelSchematic}
\end{figure}

The core of the Photo3 model, given in Section \ref{sec:modelcore}, includes the \citet{Farquhar1980} model for photosynthetic demand, an optimal control model for stomatal conductance, and a model of the soil-plant-atmosphere continuum (SPAC). In the case of C4 and CAM photosynthesis, this core is coupled with a model for carbon fixation via PEPC, which is either spatially (C4) or temporally (CAM) separated from the Rubisco-mediated Calvin cycle. The SPAC model is a simple resistor-capacitor type model of the soil-plant-atmosphere continuum (e.g. \citet{Jones1992}), and has an option to include plant water storage, which is an important feature in many succulent CAM species \citep{Nobel1988}. Given solar radiation, specific humidity, and temperature, the model estimates carbon assimilation and transpiration, as well as other variables of interest (see Figure \ref{fig:modelSchematic}).

\begin{figure}
	\includegraphics[width=15 cm]{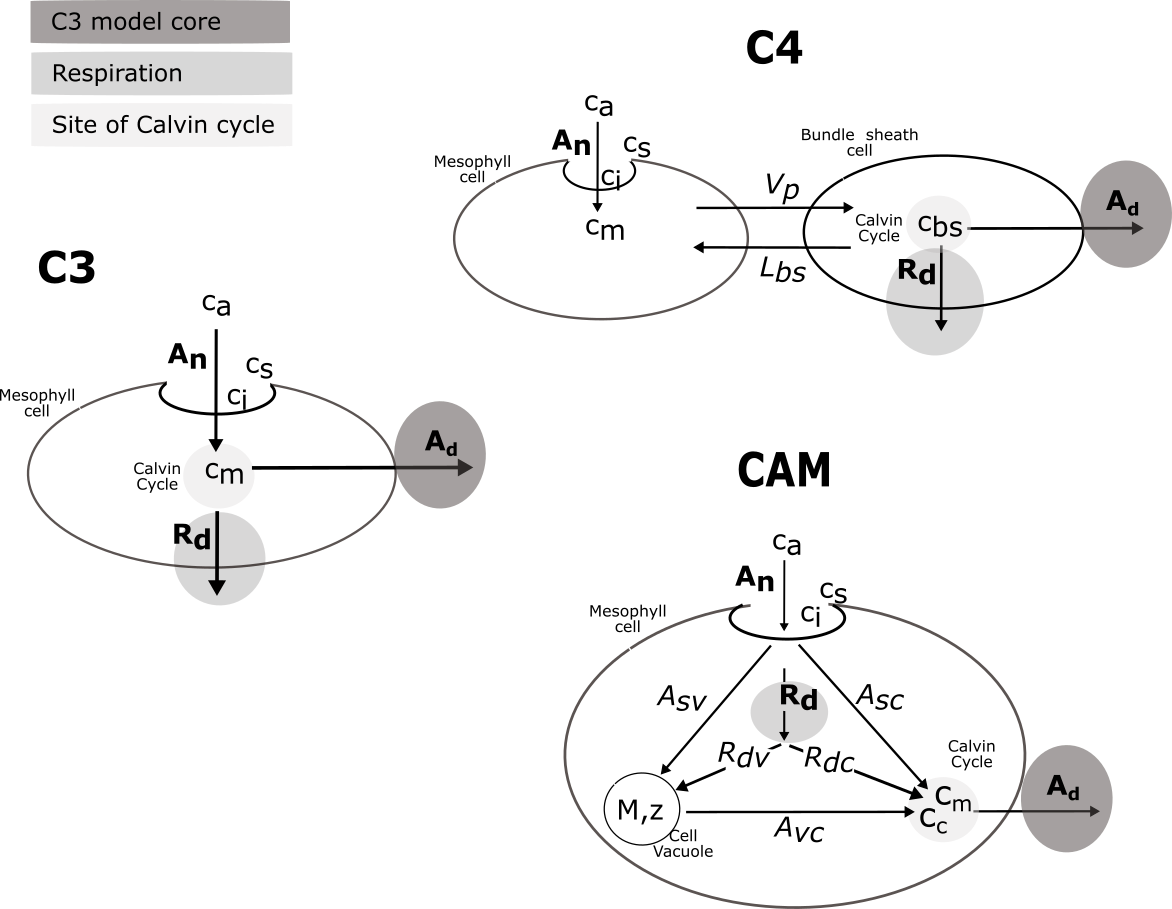}
	\caption{{\bf Representation of carbon fluxes.} The C3, C4, and CAM photosynthesis models are based on a common photosynthetic core with additional fluxes to capture the spatial and temporal separations of CO$_2$ uptake and fixation.}
	\label{fig:pSchematic}
\end{figure}

 The CAM model (formulated in Section \ref{sec:modelCAM}) includes all of the features of the core model and adds a representation of the carbon concentrating mechanism. Based on \citet{Bartlett2014}, the diurnal rhythm of malic acid production and release is modeled through the addition of a cell vacuole, characterized by $M$, the malic acid content, and $z$, the circadian order variable, which represents the overall effect of gene expression, enzyme activity and/or the vacuole tonoplast in controlling the circadian rhythm. Depending on the values of $M$ and $z$, CO$_2$ may either be fixed as malic acid by PEPC and later decarboxylated and released to the Calvin cycle, or it may be directly passed to the Calvin cycle and fixed via Rubisco. To account for the CAM idling process, dark respiration stays internal to the cell and is either passed to the cell vacuole or to the Calvin cycle, depending on the light level. These differences, described in Figure \ref{fig:pSchematic}, allow the model to capture the unique diurnal rhythm of CAM carbon fluxes \citep{Bartlett2014, Hartzell2015}. Like the CAM model, the C4 model, presented in Section \ref{sec:modelC4}, also builds on the core Farquhar model. After the initial fixation of CO$_2$ into C4 acids via PEPC, which occurs in the mesophyll cell, the C4 acids are decarboxylated and the CO$_2$ is passed to the bundle sheath cell where it is concentrated. Within the bundle sheath cells, the Rubsico-mediated Calvin cycle fixes carbon into sugar (this process is represented using the \citet{Farquhar1980} model with the new, elevated CO$_2$ concentration $c_{bs}$). A graphical description of these carbon fluxes is given in Figure \ref{fig:pSchematic}. 

\subsection{General photosynthetic relations}\label{sec:modelcore}

The net carbon uptake is modelled as a steady-state Fickian diffusion through the stomata, i.e.,
\begin{equation}
A_n = g_{s,CO_2}(c_s-c_m),
\end{equation}
where $g_{s,CO_2}$ is the stomatal conductance to CO$_2$, $c_s$ is the concentration of CO$_2$ at the leaf surface, and $c_m$ is the concentration of CO$_2$ inside the mesophyll cytosol. The stomatal conductance is assumed to scale with the square root of the vapor pressure deficit following stomatal optimization theory, an approach which agrees well with accepted empirical models \citep{Oren1999, Hari2000, Katul2009, Medlyn2011}, and can be represented as
\begin{equation}\label{eq:gms}
g_{s,CO_2} = \frac{a_1 A_n}{c_s \sqrt{D}},
\end{equation}
where $a_1$ is an empirical constant which is adjusted to account for observed differences in the $c_m/c_a$ ratio among photosynthetic types as described in \citet{Jones1992} and $D$ is the vapor pressure deficit.

The net photosynthetic demand for CO$_2$, $A_d$, is modelled according to \citet{Farquhar1980} with an adjustment to account for plant water stress, i.e.,
\begin{equation}\label{eq:Ad}
A_d = A_{\phi,c_x,T_l}(\phi, c_x, T_l) \times f_{\psi_l}(\psi_l).
\end{equation}
Due to the lack of temporal separation of CO$_2$ uptake and assimilation in C3 and C4 photosynthesis, the net carbon uptake $A_n$ is equal to the net photosynthetic demand $A_d$ for these photosynthetic types. The relationship between $A_n$ and $A_d$ for CAM includes a temporal separation and is described in Section \ref{sec:modelCAM}. The carbon demand $A_{\phi,c_x,T_l}(\phi, c_x, T_l)$ is given by
 \begin{equation}\label{eq:AphiciTl}
A_{\phi,c_x,T_l}(\phi, c_x, T_l) = \min{(A_c(c_x, T_l), A_q(\phi, c_x, T_l))},
 \end{equation}
where $A_c(c_x, T_l)$ is the Rubisco-limited photosynthetic rate, $A_q(\phi, c_x, T_l)$ is the light-limited photosynthetic rate, $T_l$ is the leaf temperature, $c_x$ is the CO$_2$ concentration at the site of the Calvin cycle, and $\phi$ is the incoming solar radiation (see Appendix \ref{sec:AppendixPhoto} for details). The relevant CO$_2$ concentration $c_x$ varies based on the photosynthetic type. For C3 plants, $c_x$ is the mesophyll concentration $c_m$; for C4 plants, it is the bundle sheath concentration $c_{bs}$; and for CAM plants, it is either $c_m$ or the corrected mesophyll CO$_2$ concentration $c_c$ (when malic acid is being decarboxylated, the mesophyll CO$_2$ concentration is corrected to account for the elevated CO$_2$ concentration as $c_c = c_m(c_a, D) + c_o f_C(z,M)$ \citep{Bartlett2014}). 

The effects of water stress reduce the photosynthetic demand according to a `vulnerability' function of the leaf water potential, $f_{\psi_l}(\psi_l)$, here represented for simplicity as a piecewise linear function which decreases between the point of onset of water stress, $\psi_{lA1}$, and the point of stomatal closure $\psi_{lA0}$, i.e.,
\begin{equation}\label{eq:Apsi}
f_{\psi_l}(\psi_l)=\left\{
\begin{array}{l l}
\text{0,} & \quad \psi_l < \psi_{lA0} \\
\frac{(\psi_l-\psi_{lA0})}{(\psi_{lA1}-\psi_{lA0})}, & \quad \psi_{lA0} < \psi_l \leq \psi_{lA1} \ \\
\text{1,} & \quad \psi_l > \psi_{lA1}, \ \\
\end{array} \right.
\end{equation}
where the leaf water potentials $\psi_{lA1}$ and $\psi_{lA0}$ are species-dependent (see Table \ref{tab:photoPar}). This piecewise function, also used in \citet{Daly2004}, provides results similar to other response functions commonly used in describing plant response to water stress, such as sigmoidal curves \citep{Sperry2002, Vico2008}, but has the advantage of providing increased numerical stability in the algorithm presented here, which involves solving relatively complex equations for the leaf water potential.

\begin{sidewaystable}[ht]
	\caption{Plant Photosynthetic Parameters\label{tab:photoPar}}
	\begin{threeparttable}
		\begin{tabular}{l l l l l l}
			\hline\hline
			Parameter &\textit{O. ficus-indica} &\textit{S. bicolor} &\textit{T. aestivum} &Units &Description\\
			\hline
			$a_1$ &2.08\tnote{a, b} &1.73\tnote{a, b} &3.46\tnote{a} &&Stomatal conductance coefficient, Eq. (\ref{eq:gms})\\
			$K_{c0}$ & 302\tnote{a} &302 &302 &$\mu$mol mol$^{-1}$ &Michaelis constant for CO$_2$ at $T_o$\\
			$K_{o0}$ & 256\tnote{a} &256 &256 &mmol mol$^{-1}$ &Michaelis constant for O$_2$ at $T_o$\\
			$H_{kc}$ & 59430\tnote{a} &59430 &59430& J mol$^{-1}$ &Activation energy for Kc\\
			$H_{ko}$ & 36000\tnote{a} &36000 &36000 & J mol$^{-1}$ &Activation energy for Ko\\
			$R_{d0}$ &0.32\tnote{a} &0.32 & 4.93\tnote{c} &$\mu$mol m$^{-2}$ s$^{-1}$ &Standard dark respiration at 25 C\\
			$H_{kR}$ & 53000\tnote{a}& 53000& 53000 & J mol$^{-1}$ &Activation energy for Rd\\
			$\kappa_2$ & 0.3\tnote{d}& 0.3& 0.3 & mol CO$_2$ mol$^{-1}$ photons &Quantum yield of photosynthesis, Eq. (\ref{eq:radiation})\\
			$V_{c,max0}$ &13\tnote{e} &39\tnote{f}& 107.4\tnote{c} & $\mu$mol m$^{-2}$ s$^{-1}$ &Maximum carboxylation rate\\
			$H_{aV}$ & 72000\tnote{g} &72000 &62000 & J mol$^{-1}$ &Activation energy for $V_{c,max}$\\
			$H_{dV}$ & 200000\tnote{g} & 200000\tnote{g} &202900\tnote{h}& J mol$^{-1}$&Deactivation energy for $V_{c,max}$\\
			$J_{max0}$ & 26\tnote{i} & 180\tnote{f} &184.9\tnote{c} &$\mu$mol m$^{-2}$ s$^{-1}$ & Electron transport rate, Eq. (\ref{eq:Jmax})\\
			$H_{aJ}$ & 50000\tnote{g} &50000 &50000 &J mol$^{-1}$ &Activation energy for $J_{max}$\\
			$H_{dJ}$ & 200000\tnote{g} &200000 &200000&J mol$^{-1}$ &Deactivation Energy for $J_{max}$\\
			$o_i$ & 0.209\tnote{a} & 0.209& 0.209&mol mol$^{-1}$ &Oxygen concentration\\
			$S_{vC}$ & 649\tnote{g} &649 &649&J mol$^{-1}$ &Entropy term for carboxylation\\
			$S_{vQ}$ & 646\tnote{g} &646 &646 &J mol$^{-1}$ &Entropy term for electron transport\\
			$T_o$ & 293.2\tnote{a} & 293.2& 293.2&K &Reference temperature\\
			$\Gamma_o$ & 34.6\tnote{a} & 34.6& 34.6&$\mu$mol mol$^{-1}$ &CO$_2$ compensation point at $T_o$, Eq. (\ref{eq:compensation})\\
			$\Gamma_1$ & 0.0451\tnote{a} & 0.0451& 0.0451&K$^{-1}$ & Eq. (\ref{eq:compensation})\\
			$\Gamma_2$ & 0.000347\tnote{a} & 0.000347& 0.000347&K$^{-2}$& Eq. (\ref{eq:compensation})\\
			$\psi_{lA1}$ & -0.5\tnote{j}& -0.5\tnote{k}& -0.7\tnote{l} & MPa &Onset of plant water stress, Eq. (\ref{eq:Apsi})\\
			$\psi_{lAO}$ & -3\tnote{j}& -1.8\tnote{k}& -2\tnote{l} & MPa &Point of maximum plant water stress, Eq. (\ref{eq:Apsi})\\
			\hline
		\end{tabular}
		\begin{tablenotes}
			\item[a] Based on \citet{Leuning1995}
			\item[b] Based on \citet{Jones1992}
			\item[c] Based on \citet{Sun2015}
			\item[d] Based on \citet{Long1993}
			\item[e] Based on \citet{Nobel1983a, Cui1993, Pimienta-Barrios2000}
			\item[f] Based on \citet{Collatz1992}
			\item[g] Based on \citet{Kattge2007}
			\item[h] Based on \citet{Daly2004}
			\item[i] Twice $V_{c, max}$, as suggested by \citet{Kattge2007}
			\item[j] Based on \citet{Bartlett2014}
			\item[k] Based on \citet{Contour-Ansel1996}
			\item[l] Based on \citet{Siddique2000}
		\end{tablenotes}
	\end{threeparttable}
\end{sidewaystable}

Since the dark respiration, $R_d$, is typically a small fraction of the carbon assimilation, its inclusion is optional except in the case of CAM where it is an important component of CAM idling behavior during periods of extended stomatal closure. Here it is represented as a temperature-dependent process according to a modified Arrhenius equation (see Appendix \ref{sec:AppendixPhoto}).

\subsection{CAM-specific relations}\label{sec:modelCAM}
In the CAM model, a carbon pool is added to represent malic acid uptake and release from the cell vacuole (see Figure \ref{fig:pSchematic}). The net flux of carbon through the stomata, $A_n$, is comprised of a flux from the stomata to the Calvin cycle, $A_{sc}$, and one to the cell vacuole, $A_{sv}$, i.e.,
\begin{equation}
A_n = A_{sc}(\phi, c_m, T_l, \psi_l, z, M) + A_{sv}(T_l, \psi_l, z, M),
\end{equation} 
where $z$ and $M$ are the circadian rhythm order variable and the malic acid concentration, respectively (described below), and the fluxes $A_{sc}$ and $A_{sv}$ are related to the carbon demand $A_d$ through the circadian state as described in Appendix \ref{sec:AppendixCAM}. The dark respiration flux, $R_d$, is likewise divided into a flux to the vacuole, $R_{dv}$, and one to the Calvin cycle, $R_{dc}$. Carbon is stored in the cell vacuole as malic acid and is released from the vacuole as the flux $A_{vc}$. The diurnal cycle of uptake and release from the vacuole, which governs the fluxes to and from the vacuole, is represented by a pair of balance equations for $M$, the malic acid concentration, and $z$, the circadian rhythm order. The balance equation for the malic acid concentration is given by
\begin{equation}
L_M \frac{dM}{dt} = A_{sv}(T_l, \psi_l, z, M) + R_{dv}(\phi, T_l) - A_{vc}(\phi, c_c, T_l, z, M),
\end{equation}
where $L_M$ is the ratio of malic acid storage volume to the carbon flux surface area, while the circadian rhythm order is given by
\begin{equation}
t_r \frac{dz}{dt} = \frac{M-M_E(z, T_l)}{M_{max}},
\end{equation}
where $t_r$ is the relaxation time, $M_E$ is the equilibrium concentration of malic acid, and $M_{max}$ is the maximum malic acid concentration. The malic acid equilibrium concentration is given by
\begin{equation}\label{eq:ME}
M_E(z, T_l) =
\left\{
\begin{array}{l l}
M_{max}\left[\left(\frac{T_H - T_l}{T_H-T_L}+1\right)c_1[\beta(z-\mu)]^3 - \frac{T_H - T_l}{T_H-T_L}[\beta(z-\mu)-c_2] + [1-f_0(z)]\right], & \quad \phi \leq 0 \ \\
M_{max}\left[\left(\frac{T_H - T_l}{T_H-T_L}+1\right)c_1[\beta(z-\mu)]^3 - \frac{T_H - T_l}{T_H-T_L}[\beta(z-\mu)-c_2]\right], & \quad \phi > 0 \ \\
\end{array} \right.
\end{equation}
where $T_H$ and $T_L$ are the high and low temperature values for the circadian oscillation, and $c_1$, $c_2$, $\mu$, and $\beta$ are circadian oscillator constants included in Table \ref{tab:CAMPar}. This formulation follows \citet{Bartlett2014} with the addition of the term $1-f_0(z)$ which synchronizes the circadian rhythm with the prevailing light cycle (this new term increases $M_E$ at high $z$ values during the night, ensuring that the uptake of malic acid continues at night even if the previous day's cycle has not completely depleted the store of malic acid). The formulations of the model fluxes are given in Appendix \ref{sec:AppendixCAM}. These adjustments to the functions given in \citet{Bartlett2014} improve model robustness to highly variable environmental inputs. 

\begin{table}
	\begin{threeparttable}
		\caption{CAM photosynthetic parameters (based on \textit{Opuntia ficus-indica})}\label{tab:CAMPar}
		\begin{tabular}{l l l l}
			\hline\hline 
			Constant & Value&Units &Description\ \\ [0.5ex] 
			\hline 
			$c_1$ & $0.365\tnote{a}$ &&Circadian oscillator constant\\
			$c_2$ & $0.55\tnote{a}$ &&Circadian oscillator constant\\
			$c_3$ & $10\tnote{a}$ &&Circadian oscillator constant\\
			$\mu$ & $0.5\tnote{a}$ &&Circadian oscillator constant\\
			$\beta$ & $2.764\tnote{a}$ &&Circadian oscillator constant\\
			$M_{max}$ &190\tnote{a}& mol m$^{-3}$ & Maximum malic acid concentration \\
			$A_{m,max}$ &13.5\tnote{a} & $\mu$mol m$^{-2}s^{-1}$ & Maximum rate of malic acid storage flux \\
			$t_r$ & 90\tnote{a} & min & Relaxation time\\
			$\alpha_1$ &1/100\tnote{a} &  & \\
			$\alpha_2$ &1/7\tnote{a} &  & \\
			$k$ &0.003\tnote{a} &  & \\
			$T_{opt}$ &288.65\tnote{a} & K & \\
			$T_{H}$ &302.65\tnote{a} & K & High temperature\\
			$T_{L}$ &283.15\tnote{a} & K & Low temperature\\
			$c_{o}$ & 3000\tnote{a} & $\mu$mol mol$^{-1}$ & Parameter for decarboxylation of malic acid\\
			[1ex] 
			\hline 
		\end{tabular}
		\begin{tablenotes}
			\item[a] Based on \citep{Bartlett2014}.
			\item[b] Empirical constant added to increase model robustness
		\end{tablenotes}
	\end{threeparttable}
\end{table}

\subsection{C4-specific relations}\label{sec:modelC4}

In the C4 plant, the influx of CO$_2$ to the bundle sheath cell is driven by the C4 pump and is modeled by a Michaelis-Menten type dependence on the mesophyll cytosol CO$_2$ concentration, $c_m$, as in \citet{VonCaemmerer2000, Vico2008} (see Figure \ref{fig:pSchematic}). The PEP regeneration rate, $V_{P}$, is bounded by the upper limit $V_{Pr}$, i.e.,
\begin{equation}\label{eq:Vp}
V_P(c_m) = \min{\left(\frac{c_m V_{P,max}}{c_m + K_p}, V_{Pr}\right)},
\end{equation}
where $V_{P,max}$ is the maximum PEP carboxylation rate and $K_p$ is the Michaelis-Menten coefficient. Leakage of CO$_2$ from the bundle sheath cell is modelled as a diffusion flux from the bundle sheath to the mesophyll cell, i.e.,
\begin{equation}\label{eq:Lbs}
L_{bs}(c_{bs}, c_m) = g_{bs}(c_{bs}-c_m),
\end{equation}
where $g_{bs}$ is the conductance between the bundle sheath and mesophyll cells and $c_{bs}$ is the CO$_2$ concentration in the bundle sheath cells.
Finally, we introduce a balance equation for the CO$_2$ fluxes into and out of the bundle sheath cell, i.e.,
\begin{equation}\label{eq:C4bal}
V_P(c_m) = A_d(\phi, c_{bs}, T_l, \psi_l) + L_{bs}(c_{bs}, c_m),
\end{equation}
and solve for the CO$_2$ concentration in the bundle sheath cell, $c_{bs}$ by combining Equation \eqref{eq:C4bal} with Equations \eqref{eq:Vp} and \eqref{eq:Lbs}. The carbon assimilation is then calculated according to Equation \eqref{eq:Ad}. Parameters for the C4 model are included in Table \ref{tab:C4Par}.

\begin{table}[ht]
	\caption{C4 Photosynthetic Parameters (based on \textit{Sorghum bicolor})}\label{tab:C4Par}
	\begin{threeparttable}
		\begin{tabular}{l l l l l l}
			\hline\hline
			Parameter &Value &Units &Description\\
			\hline
			$g_{bs}$ &.013\tnote{a}  &mol m$^{-2}$ s$^{-1}$&Conductance between bundle sheath and mesophyll\\
			$V_{P,max}$ &120\tnote{b} &$\mu$mol/m$^{-2}$ s$^{-1}$ &Maximum PEP carboxylation rate\\
			$Vpr$ & 80\tnote{c} &$\mu$mol/m$^{-2}$ s$^{-1}$ &PEP regeneration rate\\
			$Kp$ & 80\tnote{c} &$\mu$mol mol$^{-1}$ &Michaelis-Menten coefficient for C4\\
			\hline
		\end{tabular}
		\begin{tablenotes}
			\item[a] Based on \citet{Vico2008}
			\item[b] Based on \citep{Jones1992, Vico2008}
			\item[c] Based on \citep{VonCaemmerer2000, Vico2008}
		\end{tablenotes}
	\end{threeparttable}
\end{table}

\subsection{Plant hydraulics and capacitance}

The transpiration flux, $E$, is driven by the difference between the specific humidity internal to the leaf and that of the atmosphere, i.e.,
\begin{equation}\label{eq:transair}
E = g_{sa}\rho/\rho_w[q_i(T_l, \psi_l) - q_a],
\end{equation}
where $g_{sa}$ is the series of the atmospheric conductance and the combined stomatal-cuticular conductance (see Appendix \ref{sec:AppendixHydro}), $\rho$ is the density of air, $\rho_w$ is the density of water, $q_i$ is the specific humidity internal to the leaf, and $q_a$ is the specific humidity of the atmosphere. At the same time, the transpiration flux must be equal to the flux of water through the plant, i.e.,
\begin{equation}\label{eq:transplant}
E = g_{srp}(\psi_s-\psi_l),
\end{equation}
where $g_{srp}$ is the series of hte soil-root and plant conductances (see Appendix \ref{sec:AppendixHydro}), $\psi_s$ is the soil water potential, and $\psi_l$ is the leaf water potential. These relations are joined by the equation for energy balance, which equates the incoming heat flux to the outgoing sensible and latent heat fluxes, i.e.,
\begin{equation}\label{eq:energybal}
\phi = g_a \rho c_p (T_l-T_a) + \lambda_w\rho_w E,
\end{equation}
where $\phi$ is the incoming solar radiation, $g_a$ is the atmospheric conductance, $c_p$ is the specific heat of air, $T_a$ is the atmospheric temperature, and $\lambda_w$ is the latent heat of vaporization. Equations \eqref{eq:transair}, \eqref{eq:transplant} and \eqref{eq:energybal} are solved simultaneously for the three unknowns: the transpiration $E$, the leaf temperature $T_l$, and the leaf water potential $\psi_l$.

\begin{figure}
	\centering
	\includegraphics[width=8 cm]{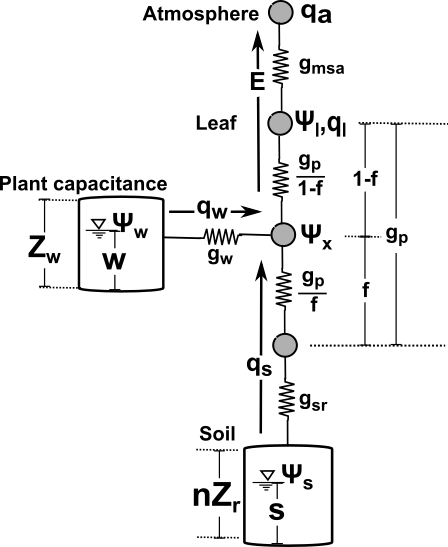}
	\caption{{\bf Representation of hydraulic fluxes.} The involved hydraulic fluxes are calculated using a resistor-capacitor model of the soil-plant-atmosphere continuum with optional plant water storage.}
	\label{fig:Cap}
\end{figure}

Plant water storage is included as an option in the model. While it is generally negligible for most C3 and C4 crops, plant water storage significantly affects water stress and carbon assimilation for plants with a substantial water storage capacity, including most CAM plants. Therefore, we include plant water storage when modeling CAM plants in this study, but not when modeling C3 and C4 plants. Plant water storage is represented in this model as a simple capacitor located at a height which is a fraction $f$ of the total plant height as in \citet{Hartzell2017} (see Figure \ref{fig:Cap} and Appendix \ref{sec:AppendixHydro}). Using this scheme, the change in the plant relative water content $w$ is given by the balance equation
\begin{equation}
LAI Z_w \frac{dw}{dt} = q_s(s, w) - E,
\end{equation}
where $Z_w$ is the total available water storage depth of the plant on a leaf area basis, $LAI$ is the leaf area index, and $s$ is the soil moisture. The transpiration flux $E$ is now given by the sum of the fluxes from the soil, $q_s$, and the plant water storage, $q_w$, such that
\begin{equation}\label{eq:psi_x}
E = q_s + q_w = g_{srpf}(\psi_s-\psi_x)+g_w LAI (\psi_w-\psi_x),
\end{equation}
where $\psi_x$ is the water potential at the storage connection node, $g_{srfp}$ is the conductance between the soil and storage connection node, $g_w$ is the conductance between the site of water storage and the storage connection node, and $\psi_w$ is the water potential of the water storage tissue (see Appendix \ref{sec:AppendixHydro} and Table \ref{tab:hydPar} for details). The addition of plant water storage adds a fourth unknown ($\psi_x$) to the water balance, which may now be formulated as described in Appendix \ref{sec:AppendixHydro}.

\begin{sidewaystable}[ht]
	\caption{Plant Hydraulic Parameters \label{tab:hydPar}}
	\begin{threeparttable}
		\begin{tabular}{l l l l l l}
			\hline\hline
			Parameter &\textit{O. ficus-indica} &\textit{S. bicolor} &\textit{T. aestivum} &Units &Description\\
			\hline
			$g_{pmax}$ & 0.04\tnote{a} &0.13\tnote{i} &11.7\tnote{d} & $\mu$m MPa$^{-1}$ s$^{-1}$ &Maximum xylem conductance per unit leaf area\\
			$LAI$ & 3\tnote{b} &5\tnote{j} &5\tnote{o}&m$^2$ m$^{-2}$& Leaf area per unit ground area \\
			$Z_r$&0.1\tnote{c} & 0.5\tnote{k} &0.75\tnote{p}&m&Mean rooting depth\\
			$RAI_w$ & 3\tnote{c} &5.6\tnote{l} &5.6\tnote{l}&m$^2$ m$^{-2}$ &Root area index under well watered conditions\\
			$d$ &8\tnote{d} &8\tnote{d} &8\tnote{d} &&Parameter accounting for root growth\\
			$\alpha_m$ & 1\tnote{e} & 2.65\tnote{l} &1.65\tnote{l} & mol m$^{-2}$ s$^{-1}$ & Ratio of g$_{m,CO2}$ to g$_{s,CO2}$\\
			$g_{cut}$ & 0\tnote{b} &0.1802\tnote{m} &0.3\tnote{q}& mm s$^{-1}$ &Cuticular conductance per unit leaf area\\
			$g_a$ & 324\tnote{f}& 61\tnote{n} & 61\tnote{n}& mm s$^{-1}$ &Atmospheric conductance per unit ground area\\
			$Z_w$ & 0.00415\tnote{g} & - & -  & m$^3$ m$^{-2}$ &Maximum depth of water stored per unit leaf area\\
			$g_{wmax}$ & 0.002\tnote{a} & - & -  & $\mu$m MPa$^{-1}$ s$^{-1}$&Maximum conductance between stored water and transport pathway, per unit leaf area\\
			$h$ & 2\tnote{d} & - & - & &Parameter for Eq. (\ref{eq:gp})\\
			$j$ & 2\tnote{d} & - & - & &Parameter for Eq. (\ref{eq:gp})\\
			$m$ & 4\tnote{h} & - & - & &Parameter for Eq. (\ref{eq:gw})\\
			$c$ & 0.27\tnote{g} & - & -  &MPa$^{-1}$& Intrinsic plant hydraulic capacitance, Eq. (\ref{eq:psiw})\\
			$f$ & 0.5 & - & - &&Fractional height of hydraulic capacitance\\
			
			\hline
		\end{tabular}
		\begin{tablenotes}
			\item[a] Estimated from a succulent CAM species based on \citet{Bartlett2014}
			\item[b] Based on \citet{Nobel1988}
			\item[c] Based on \citet{Snyman2005}
			\item[d] Based on \citet{Daly2004}
			\item[e] Based on \citet{Flexas2008}
			\item[f] Based on \citet{Jones1992} for a windspeed of 2 m/s at 2 m altitude, with a plant height of 2 m.
			\item[g] Based on \citet{Goldstein1991a}
			\item[h] Based on values from \citet{Waring1978, Carlson1991}
			\item[i] Based on \citet{Kocacinar2003}
			\item[j] Based on \citet{Olufayo1996}
			\item[k] Based on \citet{Bremner1986}
			\item[l] Based on \citet{Vico2008}
			\item[m] Based on \citet{Muchow1989}
			\item[n] Based on \citet{Jones1992} for a windspeed of 2 m/s at 2 m altitude, with a plant height of 1 m.
			\item[o] Based on \citet{Lunagaria2006}
			\item[p] Based on \citet{Bandyopadhyay2003}
			\item[q] Based on \citet{Kerstiens1996}
		\end{tablenotes}
	\end{threeparttable}
\end{sidewaystable}

The soil moisture may either be provided as a model input or determined through the balance equation,
\begin{equation}\label{eq:dsdt}
nZ_r\frac{ds}{dt} = -q_s(s,w) - L(s) - E_v(s) + R(t),
\end{equation}
where $n$ is the soil porosity, $Z_r$ is the rooting depth and $s$ is the volumetric soil moisture averaged over the rooting depth. Total losses from the soil are due to plant water uptake $q_s(s, w)$, leakage loss $L(s)$, and evaporation $E_v(s)$ (see \citet{Rodriguez2004} for details). In Equation \eqref{eq:dsdt}, the rainfall $R(t)$ may be specified either as a model input or may be generated within the model as a stochastic process which requires the mean rainfall depth and frequency as input parameters. Currently a range of soil types are represented by the four soil options included in the model: loamy sand, sandy loam, loam, and clay. Soil parameters are included in Table \ref{tab:soilPar}.

\begin{table}[ht]
	\caption{Soil Parameters\tnote{a} \label{tab:soilPar}}
	\begin{threeparttable}
		\begin{tabular}{l l l l l}
			\hline\hline
			Parameter &Loamy sand &Sandy loam &Loam &Clay\\
			\hline
			$K_s$ (cm d$^{-1}$) &100 &80  &20 &1\\
			$\overline{\psi}_s$ (MPa)&-1.7$\cdot$10$^{-4}$ & -7.0$\cdot$10$^{-4}$ &-1.43$\cdot$10$^{-3}$ & -1.82$\cdot$10$^{-3}$\\
			$b$ &4.38 &4.9 &5.39 &11.4\\
			$n$ &0.42 & 0.43  &0.45 &0.5\\
			$s_h$ &0.08 & 0.14  &0.19 &0.47\\
			\hline
		\end{tabular}
		\begin{tablenotes}
			\item[a]Parameters from \citet{Rodriguez2004}
		\end{tablenotes}
	\end{threeparttable}
\end{table}

\subsection{Model Implementation}

The model requires inputs of environmental conditions including solar radiation, air temperature, specific humidity, soil moisture, and soil type. Because of the strong dependence of CAM photosynthesis on variability in environmental conditions at the sub-daily scale, the model operates with a 30-minute timestep. Solar radiation, specific humidity, and air temperature data with an hourly timescale may be interpolated to give values at each model timestep. Alternatively, values for these variables may be generated internally to the model using a built-in boundary-layer simulation following the approach presented in \citet{Daly2004}. The model is currently parameterized with hydraulic and photosynthetic properties for three representative species, \textit{Triticum aestivum} (C3), \textit{Sorghum bicolor} (C4), and \textit{Opuntia ficus-indica} (CAM) (see Tables \ref{tab:photoPar}, \ref{tab:C4Par}, \ref{tab:CAMPar}, and \ref{tab:hydPar}); these species were selected because they are among the most well-studied and economically important species of each photosynthetic group \citep{Leff2004, Paterson2008}. These properties are meant to represent plants at the mature stage in the growing season, and are assumed to be approximately constant over the model duration.

\subsection{Model validation and testing}

\subsubsection{Model validation}
The model was validated using data collected under both well-watered and droughted conditions for the three representative crops (see Section \ref{sec:validation}). Model results under well-watered conditions for \textit{Opuntia ficus-indica} were compared to results from a 24-hour laboratory experiment undertaken by \citet{Nobel1983a} with 12 hours of light and 12 hours of darkness. In this simulation, the day period was characterized by a solar radiation of 244 W/$m^2$, a temperature of 25 C, and a relative humidity of 40\%, while the night period was characterized by a temperature of 15 C and a relative humidity of 60\% according to the conditions present in the laboratory experiment. To facilitate comparisons of model performance with experimental data for \textit{S. bicolor} and \textit{T. aestivum}, the model was forced with typical non-limiting laboratory conditions of 12 h light:12 h darkness with a photosynthetic photon flux density (PPFD) of 1800 $\mu$mol/m$^2$/s during the light period, a constant temperature of 26 C, constant relative humidity of 80\%, and 0.7 volumetric soil moisture in loam soil. To enable comparison of model results with data for \textit{O. ficus-indica}, the model was run for a 40 day drydown in loamy sand with solar radiation, temperature, and relative humidity obtained from the National Solar Radiation Database (NSRDB) \citep{NREL2017} on March 17, 2015 at a weather station nearby the study location in Til Til, Chile. Results were compared with data from \citet{Acevedo1983} describing carbon assimilation and stomatal conductance of  under water-stressed conditions,

\subsubsection{Diurnal dynamics}
The diurnal dynamics of the three photosynthetic types were compared under typical, well-watered, growing conditions (see Section \ref{sec:diurnal}). The model was run using meteorological data from Temple, TX on April 1, 2015, imposing a soil moisture of 0.5 and a soil type of sandy loam. The solar radiation, air temperature, and specific humidity for the site were obtained from the NSRDB data viewer \citep{NREL2017}. 

\subsubsection{Long-term performance under drought}
We evaluated the relative long-term performance of the three crops during a drought (Section  \ref{sec:drought}) by simulating a drydown of 40 days was simulated beginning with a volumetric soil moisture of 0.5 and using temperature, relative humidity, and solar radiation data from Temple TX from April 1, 2015 until May 10, 2015 \citep{NREL2017}. During the drydown, the soil moisture was calculated according to Eq. \eqref{eq:dsdt}.

\subsubsection{Representation of C3-CAM intermediates}
To test the ability of the model to capture C3/CAM intermediate photosynthetic types, the model was executed with various levels of maximum malic acid storage capacity, $M_{max}$ (see Section \ref{sec:CAMint}). Values tested were $M_{max}$ = 190 mol/m$^3$ (default model setting), 95 mol/m$^3$ (50\% of the default setting), and 1.9 mol/m$^3$ (1\% of the default setting). Results from these simulations were compared to a simulation run with all the photosynthetic parameters of \textit{O. ficus-indica}, but with the photosynthetic type set equal to C3 rather than CAM. In each of these simulations, plant water storage and respiration were not included. Simulations were run for sandy loam soil with a constant soil moisture of 0.5 and weather conditions found in Temple, TX on April 3, 2015 \citep{NREL2017}.

\section{Results and discussion}

\subsection{Model validation}\label{sec:validation}

\begin{figure}
	\centering
	\includegraphics[width=17 cm]{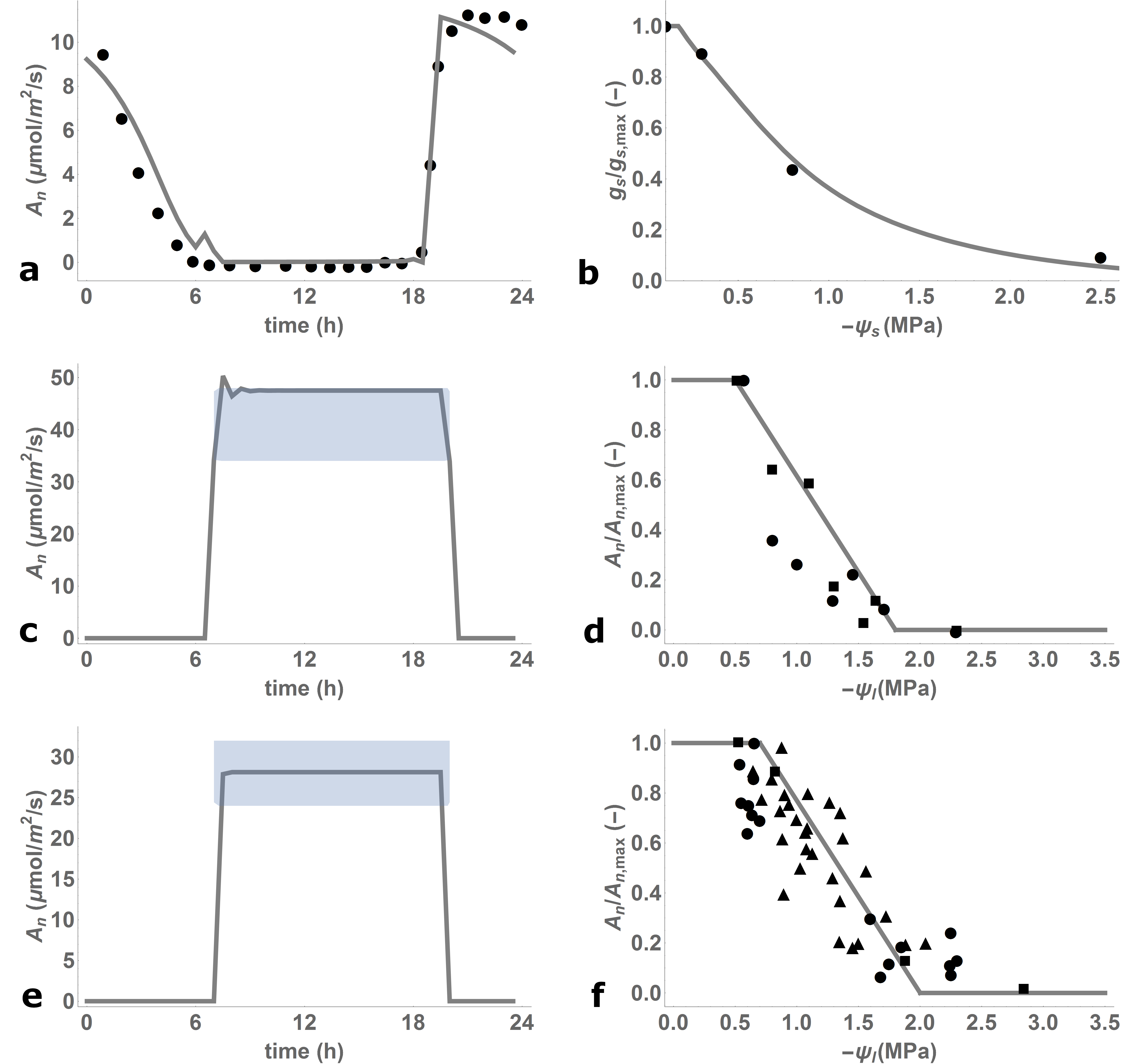}
	\caption{{\bf Comparison with experimental data.} (a) Comparison of modeled carbon assimilation, $A_n$ ($\mu$mol/m$^2$/s) for \textit{Opuntia ficus-indica} (solid line) with data from \citet{Nobel1983a} (circles). (b) Comparison of modeled decrease in stomatal conductance $g_s$ as a fraction of maximal stomatal conductance $g_{s,max}$ for \textit{O. ficus-indica} during a drydown period with data from \citet{Acevedo1983} (circles). (c) Comparison of modeled carbon assimilation as a function of maximal carbon assimilation, $A_{n,max}$, for \textit{Sorghum bicolor} with published ranges in laboratory experiments according to \citet{Peng1990, Resende2012} (gray shading). (d) Comparison of modeled decrease in carbon assimilation with leaf water potential for \textit{S. bicolor} and data from \citet{Contour-Ansel1996} for two cultivars: ICSV 1063 (circles) and MIGSOR (squares). (e) Comparison of modeled carbon assimilation for \textit{Triticum aestivum} with published ranges in laboratory experiments according to \citet{Evans1983, Martin1992} (gray shading). (f) Comparison of modeled decrease in carbon assimilation with leaf water potential and data for four cultivars of \textit{T. aestivum} (Kanchan, Sonalika, Kalyansona, and C306) from \citet{Siddique2000} (circles), data for TAM W-101 \citep{Johnson1987} (squares), and data for TAM W-101 and Sturdy \citep{Martin1992} (triangles). }
	\label{fig:Opuntia}
\end{figure}

For \textit{O. ficus-indica}, the magnitude and diurnal dynamics of both carbon assimilation (see Figure \ref{fig:Opuntia}a) and stomatal conductance (data not shown) closely match those observed in controlled laboratory experiments. Carbon assimilation reaches a maximal value of 10 $\mu$mol/m$^2$/s, in agreement with the data, and stomatal conductance reaches a maximal value of 3.0 mm/s, as compared with the observed maximal value of 2.8 mm/s. The diurnal dynamics show a relatively good fit, with a slight underestimate of carbon assimilation in the middle of the night (hours 21-1) and a slight overestimate of carbon assimilation at dawn (hour 6). At dusk, the timing of the onset of CAM carbon assimilation matches very well with the data, while the decrease in carbon assimilation and stomatal conductance at dawn is slightly slower than that observed. For \textit{S. bicolor}, the carbon assimilation rate under optimal conditions is 48 $\mu$mol/m$^2$/s, which lies within the range of published experimental ranges of 34-48 $\mu$mol/m$^2$/s \citep{Peng1990} and 40 $\mu$mol/m$^2$/s \citep{Resende2012}  (see Figure \ref{fig:Opuntia}c). For \textit{T. aestivum}, the maximal simulated carbon assimilation rate is 28 $\mu$mol/m$^2$/s, which agrees well with experimental values of 24-29 $\mu$mol/m$^2$/s \citep{Evans1983} and 32 $\mu$mol/m$^2$/s \citep{Martin1992}  (see Figure \ref{fig:Opuntia}e).

Model responses to water limitations also compare well with data for the three representative species. Figure \ref{fig:Opuntia}b shows daily maximal stomatal conductance for \textit{O. ficus-indica} at various levels of soil water potential simulated during a drydown. The daily maximal stomatal conductance decreases from a maximum under well-watered conditions to 50\% of its original value at a soil water potential of approximatley -0.7 MPa. This behavior is a good fit with field measurements of stomatal conductance taken during a progressive drydown \citep{Acevedo1983}. Finally, the model response to water limitation $f_{\psi_l}(\psi_l)$, given by Equation \eqref{eq:Apsi}, is compared with experimental data for carbon assimilation at a range of leaf water potentials for both \textit{S. bicolor} and \textit{T. aestivum}. Data shown include two cultivars of \textit{S. bicolor}: ICSV 1063 and MIGSOR (data from \citet{Contour-Ansel1996}), and several cultivars of \textit{T. aestivum}: Kanchan, Sonalika, Kalyansona, and C306 (data from \citet{Siddique2000}), TAM W-101 (data from \citep{Johnson1987}), and TAM W-101 and Sturdy (data from \citep{Martin1992}) (see Figure \ref{fig:Opuntia}d, f). The response of the model to moisture limitations generally agrees with published data for these species and provides a fit similar to that obtained through a sigmoidal function of leaf water potential.

\subsection{Diurnal dynamics}\label{sec:diurnal}

\begin{figure}
	\centering
	\includegraphics[width=17 cm]{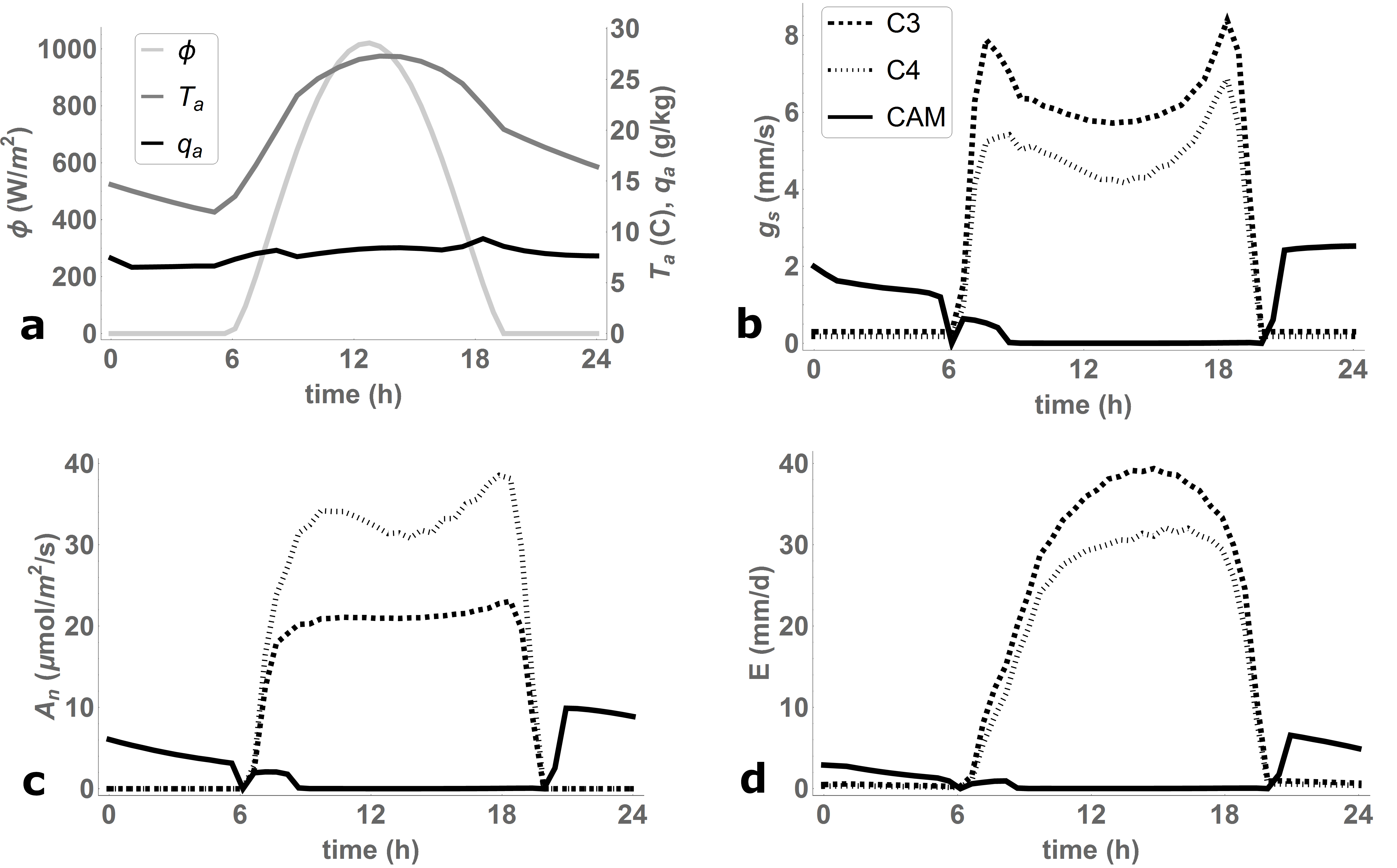}
	\caption{{\bf A comparison of the diurnal behavior of the three photosynthetic types.} (a) Model inputs of solar radiation $\phi$ (W/m$^2$), temperature $T$ (C) and specific humidity $q_a$ (g/kg) obtained using data from Temple, TX on April 30, 2015. (b) Simulated stomatal conductance to water vapor $g_s$ (mm/s) on a leaf area basis, (c) simulated carbon assimilation $A_n$ ($\mu$mol/m$^2$/s) on a leaf area basis, and (d) simulated transpiration $E$ (mm/d) on a ground area basis for  \textit{Triticum aestivum} (C3), \textit{Sorghum bicolor} (C4), and \textit{Opuntia ficus-indica} (CAM) under non-water-limited conditions (soil moisture of 0.5).}
	\label{fig:dailyPanel}
\end{figure}

Due to its detailed representation of CAM dynamics, the Photo3 model is able to compare CAM, C3, and C4 functioning at a half-hourly timescale. Figure \ref{fig:dailyPanel}a shows the solar radiation, air temperature, and specific humidity for a typical April day in Temple, TX. Model responses of carbon assimilation, transpiration, and stomatal conductance for each of the photosynthetic types are shown in Figure \ref{fig:dailyPanel}b-d. The characteristic stomatal behavior of CAM is clearly shown, with stomata opening primarily at night. A combination of a low stomatal conductance and a low nocturnal driving force for transpiration during this period result in a very low transpiration with a moderate nocturnal carbon assimilation. The carbon concentrating behavior of C4 results in a high maximum carbon assimilation, while the stomatal conductance and transpiration are slightly lower than that of C3. Under these conditions, the C4 plant shows a high productivity and a high water use efficiency, while the CAM plant shows a relatively low productivity but a very high water use efficiency. 

\subsection{Long-term performance under drought}\label{sec:drought}

\begin{figure}
	\centering
	\includegraphics[width=17 cm]{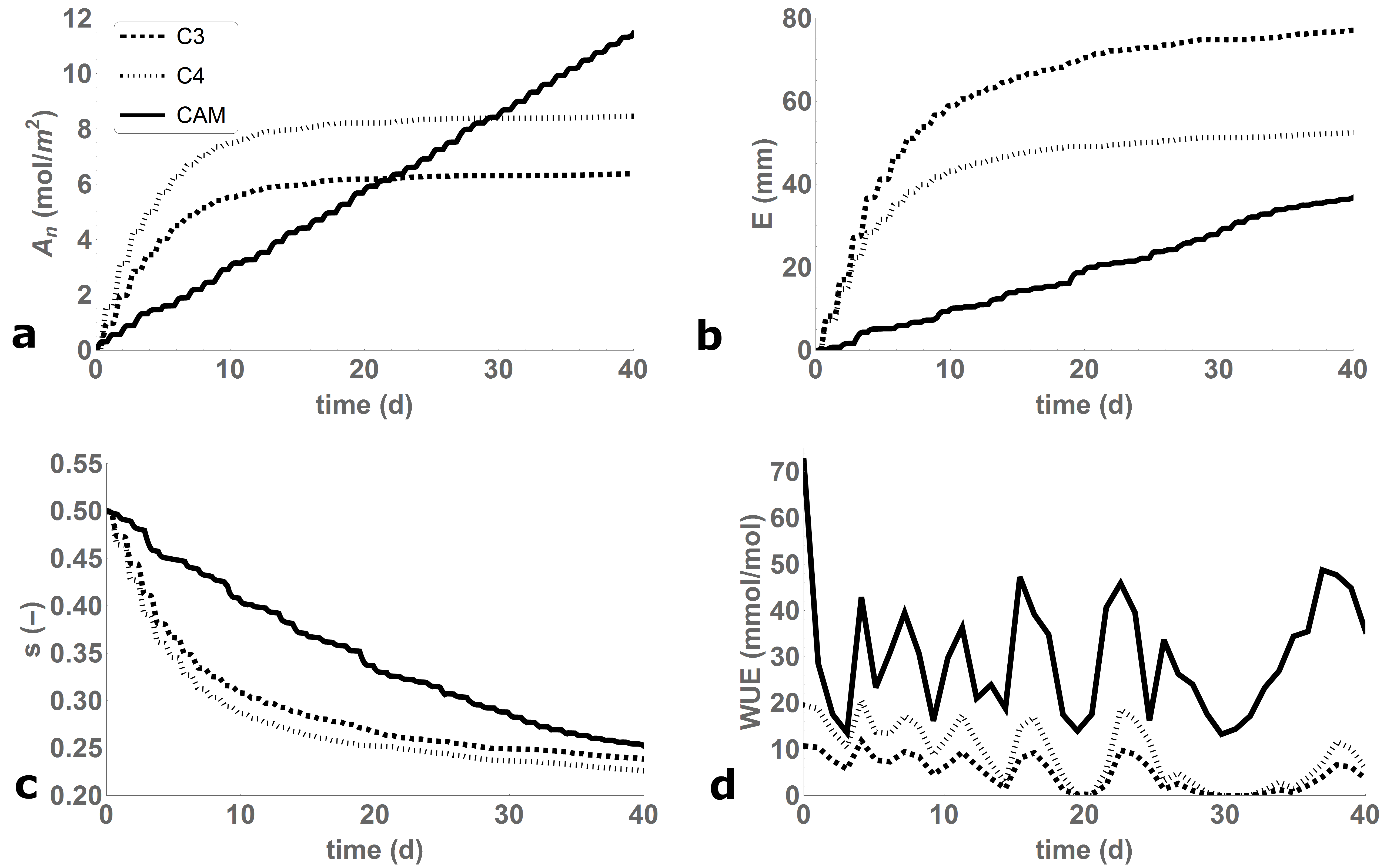}
	\caption{{\bf A comparison of the long-term performance under drought for the three photosynthetic types.} Simulated (a) cumulative carbon assimilation $A_n$ (mol/m$^2$) on a leaf area basis, (b) cumulative transpiration $E$ (mm) on a leaf area basis, (c) volumetric soil moisture $s$, and (d) instantaneous water-use efficiency $WUE$ (mmol/mol) over the course of 40 days beginning on April 1, 2015 in Temple, TX.}
	\label{fig:cumulativePanel}
\end{figure}

Model simulations of cumulative carbon assimilation and water use during a drought period are shown in Figure \ref{fig:cumulativePanel} for the three photosynthetic types.  While the C3 and C4 crops initially have high productivity, assimilating carbon at a rate two to three times that of the CAM crop, the productivity of the C3 and C4 crops undergoes a large decrease after about 8-10 days as the soil dries to below 0.3 volumetric soil moisture (Figure \ref{fig:cumulativePanel}a). Meanwhile, the CAM crop exhibits a slower, but more persistent rate of carbon gain. By day 22, the total carbon assimilation of \textit{O. ficus-indica} surpasses that of \textit{T. aestivum}, and by day 29, it surpasses that of \textit{S. bicolor}. The CAM crop also shows a much lower cumulative transpiration, by a factor of nearly five, during the early stages of the drought, while the overall water use of the C3 and C4 crops are similar during this period (see Figure \ref{fig:cumulativePanel}b). This allows the soil moisture in the CAM simulation to remain at a much higher level for the first 20 days of the simulation (see Figure \ref{fig:cumulativePanel}c). While the CAM species exhibits a lower productivity under non-water-limited conditions, it exhibits a much higher water use efficiency and its productivity persists longer under water-limited conditions. After a 40-day drydown, the CAM crop ultimately assimilates twice as much carbon as the C3 species and 50\% more than the C4 species. At the same time, its total transpiration is less than half that of the C3 species and about 70\% that of the C4 species. Depending  on the specific environmental conditions, the photosynthetic water use efficiency of CAM is two to six times higher than C3, and one to five times higher than C4 (Figure \ref{fig:cumulativePanel}d). This consistent basis of comparison, which incorporates the effects of environmental variability at both long and short timescales, allows the model user to quantify the costs and benefits of crops with different photosynthetic types in water-limited ecosystems.

\subsection{Representation of C3-CAM intermediates}\label{sec:CAMint}

CAM photosynthesis is generally not a discrete trait, rather, a spectrum of C3-CAM behavior exists in nature  \citep{Winter2015, Brautigam2017}. Metabolite fluxes similar to CAM fluxes have been shown in C3 plants, to a much smaller degree \citep{Winter2015, Brautigam2017}, and some cacti, including \textit{O. ficus-indica}, \textit{Agave deserti}, and \textit{Mesembryanthemum crystallinum} show dramatic changes in the level of CAM expression throughout their lifespan, switching from C3 to CAM photosynthesis during the process of development or in response to water stress \citep{Kluge1978, Winter1978, Jordan1979, Acevedo1983, Winter2008, Winter2011}. The consistent formulations of the three different photosynthetic pathways in the Photo3 model allows intermediate CAM-C3 behavior to be explored through the adjustment of a single model parameter. This is the first time that such an analysis has been possible in a model coupled to the soil-plant-atmosphere continuum. 

\begin{figure}
	\includegraphics[width=17 cm]{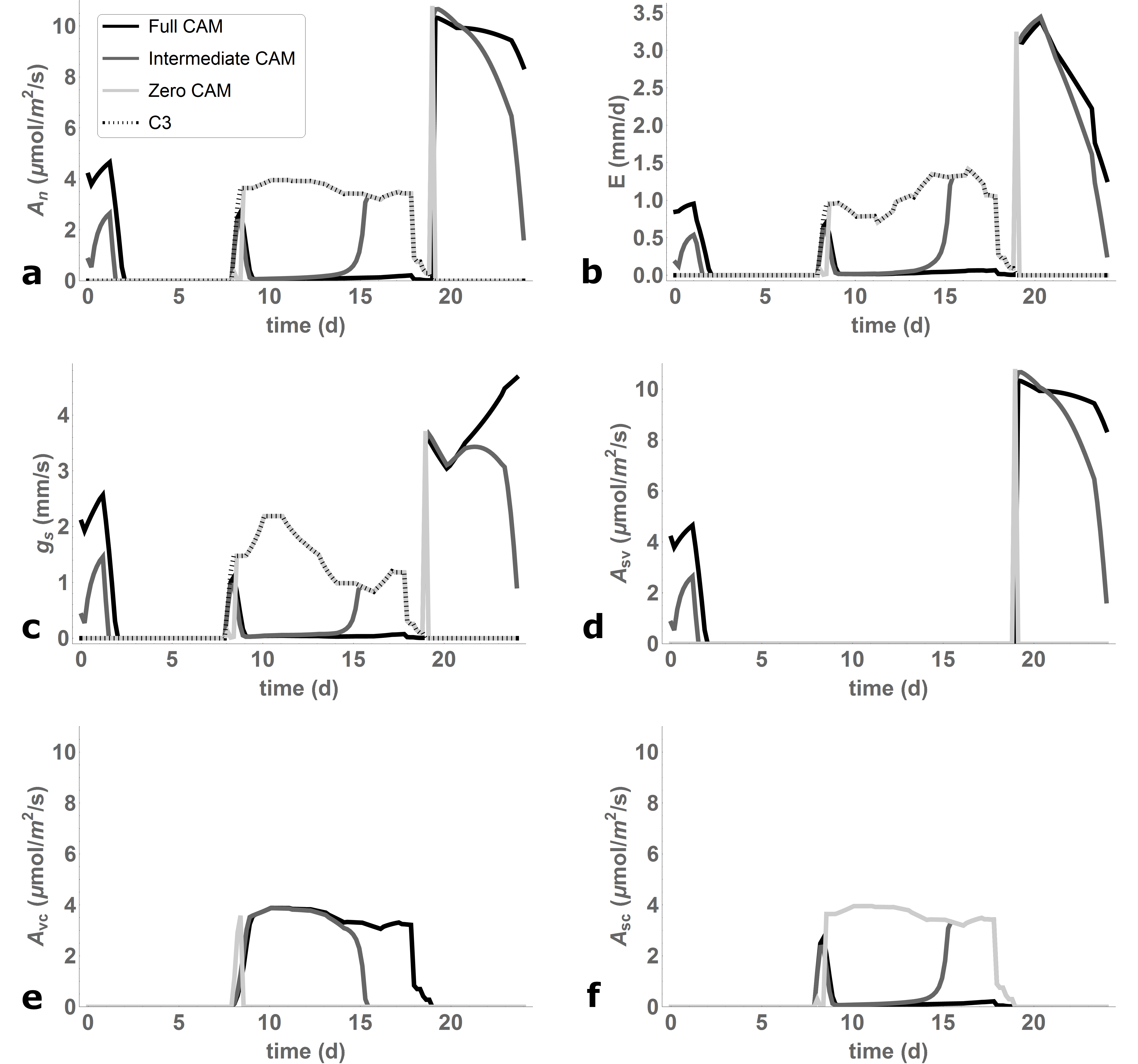}
	\caption{{\bf Simulation of CAM-C3 intermediates.} (a-c) Comparison of CAM-C3 intermediates with C3 photosynthesis. Simulated (a) carbon assimilation $A_n$ ($\mu$mol/m$^2$/s) on a leaf area basis, (b) transpiration $E$ (mm/d) on a ground area basis, and (c) stomatal conductance $g_s$ (mm/s) to water vapor on a leaf area basis. (d-f) Carbon fluxes for CAM-C3 intermediates. Fluxes (d) A$_{sv}$ from the stomata to the vacuole, (e) A$_{vc}$ from the vacuole to the Calvin cycle, and (f) A$_{sc}$ from the stomata to the Calvin cycle. All simulations are based on parameters for \textit{Opuntia ficus-indica} with a constant soil moisture of 0.5 and weather conditions found in Temple, TX on April 3, 2015.}
	\label{fig:camPanel}
\end{figure}

As CAM expression becomes stronger, vacuole size and maximum malic acid storage capacity increase. Indeed, maximum malic acid content is a typical measure of the strength of CAM expression \citep{Kluge1978}. By altering the maximum malic acid concentration $M_{max}$ in the CAM model, C3 behavior can be retrieved from the CAM model framework. These results are shown in Figure \ref{fig:camPanel}a-c, which shows carbon assimilation, transpiration, and stomatal conductance for \textit{O. ficus-indica} with varying degrees of malic acid storage capacity (full CAM expression with 100\% malic acid storage capacity, CAM intermediate with 50\% malic acid storage capacity, and CAM intermediate with 1\% malic acid storage capacity), and with C3 type photosynthesis. As the maximum malic acid concentration approaches zero, the length of nocturnal stomatal opening becomes shorter until it approaches zero in accordance with the C3 model. Meanwhile, the stomata begin to open for increasingly longer periods during the day. For very low values of maximum malic acid concentration, the stomatal conductance, carbon assimilation, and transpiration of the CAM model match those of the C3 model. 

The relevant CAM fluxes $A_{sv}$, $A_{vc}$, and $A_{sc}$ are shown in Figure \ref{fig:camPanel}d-f for various levels of CAM expression. As $M_{max}$ decreases, the fluxes $A_{sv}$ and $A_{vc}$ decrease, while the duration of $A_{sc}$ increases until it is occurring for all daylight hours as $M_{max}$ approaches zero. This behavior can be understood by referring to Appendix \ref{sec:AppendixCAM}. As $M_{max}$ decreases, the amount of malic acid stored during the night becomes smaller, and the malic acid is more quickly depleted during the day. As a result, the carbon function $f_C$, which depends on the malic acid concentration (see Equation \eqref{eq:fC}) is at a non-zero value for a briefer period after the onset of light. As $f_C$ approaches zero during the day, $A_{vc}$ approaches zero and $A_{sc}$ approaches $A_n$, yielding C3-like behavior (see Equations \eqref{eq:Avc} and \eqref{eq:Asc}). Likewise, the flux $A_{sv}$ is restricted to smaller time increments at dawn and dusk and finally approaches zero according to Equation \eqref{eq:Asv}.


\section{Conclusions}

Results of the Photo3 model allow the comparison of CAM, C3, and C4 species in a consistent framework. This detailed model is streamlined, user-friendly, and robust to a wide range of environmental conditions. Thus, it is ready to be included as a component of earth system models, crop models, and bioenery models. In each of these areas, a more detailed representation of CAM functioning can illuminate important questions. The inclusion of C4 plants in earth system models has been evaluated and has been shown to have a significant effect on land cover and on local climate conditions \citep{Cowling2007}. Thus, CAM plants, which arguably inhabit more extreme climates, may also be expected to show important local effects. A more detailed representation of CAM in crop modeling will be a useful tool for evaluating potential productivity, planting strategies, and water use strategies for CAM agriculture. The open source, modular nature of the Photo3 model allows for the addition of other features that may help to explore a wide range of research questions. In this stage of its development, Photo3 currently assumes static plant traits like leaf area index, root area index, and photosynthetic capacity, and outputs plant water use and carbon assimilation as a function of climate conditions. In the future, additional modules may be added to support the inclusion of more inputs, such as plant nutrient status and growing stage, and outputs like biomass accumulation and crop yield. As pressure increases to provide food, water, and energy to a growing population, ecohydrological modeling tools such as Photo3 will be necessary to quantitatively evaluate the tradeoffs between species with different photosynthetic strategies \citep{Porporato2015}.

The Photo3 model, for the first time, provides a physiologically based and consistent representation of the three photosynthetic types (CAM, C3, and C4) coupled to environmental conditions. This is done using a consistent model core built on the \citet{Farquhar1980} model for carbon assimilation, a model of stomatal functioning based on stomatal optimality, and a resistor-capacitor model of the soil-plant-atmosphere continuum. The model allows the user to compare expected productivity and water use efficiency of CAM plants directly with that of C3 and C4 plants under a wide range of climate conditions. The results produced here for simulations of \textit{Opuntia ficus-indica}, \textit{Triticum aestivum}, and \textit{Sorghum bicolor}, show detailed predictions of stomatal conductance, carbon assimilation, and transpiration at the daily level and also facilitate the comparison of long-term carbon assimilation and water use of the three photosynthetic types under drought conditions. Through the adjustment of a single model parameter, the model framework is also able to capture intermediate C3-CAM behavior. The open source, modular nature of the model is designed for to be user friendly and easy to couple with existing modeling efforts. Photo3 shows promise for use in a variety of research applications where models of CAM photosynthesis are currently lacking, including the prediction of CAM climate feedbacks, productivity, and biofuel potential.

\section*{Software availability}

The Photo3 software can be accessed  for free at https://samhartz.github.io/Photo3/. It was created by Samantha Hartzell, Mark Bartlett, and Amilcare Porporato (email aporpora@princeton.edu, phone 609 258 2287), and first made available in 2017. Photo3 was developed in Python 2.7 with the SciPy, NumPy, Pandas, Tkinter, and Matplotlib packages. Program size is 45.6 KB. We suggest installing a Python distribution such as Anaconda to meet the program requirements.

\section*{Acknowledgements}

This work was supported through the USDA Agricultural Research Service cooperative agreement 58-6408-3-027 and National Institute of Food and Agriculture (NIFA) grant 12110061; and National Science Foundation (NSF) grants CBET-1033467, EAR-1331846, FESD-1338694, EAR-1316258, GRFP-1106401 and the Duke WISeNet Grant DGE-1068871. We thank Marina Smalling for her useful feedback. 

\section*{Appendix}
\renewcommand{\thesubsection}{\Alph{subsection}}
\renewcommand{\theequation}{\Alph{subsection}.\arabic{equation}}
\counterwithin{equation}{subsection}

\subsection{Details of the general photosynthetic model}\label{sec:AppendixPhoto}
The Rubisco-limited rate of carbon assimilation, $A_c$, is given by
\begin{equation}
A_c(c_x, Tl)=V_{c,max}\frac{c_x-\Gamma^*(T_l)}{c_x+K_c(T_l)(1+o_i/K_o(T_l))},
\end{equation}
where $V_{c,max}$ is the maximum carboxylation rate, $c_x$ is the relevant CO$_2$ concentration (see Equation \eqref{eq:AphiciTl}), $K_c$ and $K_o$ are the Michaelis-Menten coefficients for CO$_2$ and O$_2$, respectively, o$_i$ is the oxygen concentration, and $\Gamma^*$ is the CO$_2$ compensation point (see Table \ref{tab:photoPar} for the model parameters). The maximum carboxylation rate, $V_{c,max}$, and the CO$_2$ compensation point, $\Gamma^*$, are given by
\begin{equation} \label{eq:Vmax}
V_{c,max}(T_l)=V_{c,max0}\frac{\exp[\frac{H_{aJ}}{RT_0}(1-\frac{T_0}{T_l})]}{1+\exp(\frac{S_{vC}T_l-H_{dJ}}{RT_l})}
\end{equation}
and
\begin{equation} \label{eq:compensation}
\Gamma^*(T_l)=\Gamma_0[1+\Gamma_1(T_l-T_0)+\Gamma_2(T_l-T_0)^2],
\end{equation}
where $R$ is the universal gas constant (J/(mol K)), $T_0$ is a reference temperature, and the remaining parameters are given in Table \ref{tab:photoPar}. The temperature dependence of the Michaelis-Menten constants $K_c$ and $K_o$ is described by a modified Arrhenius equation, i.e.,
\begin{equation}
K_x(T_l)=K_{x0}\exp\left[\frac{H_{Kx}}{RT_0}\left(1-\frac{T_0}{T_l}\right)\right].
\end{equation}

The light-limited assimilation rate, $A_q$, is given by
\begin{equation}
A_q(\phi, c_x, Tl) = \frac{J(\phi, T_l)}{4}\frac{(c_x-\Gamma^*(T_l))}{(c_x+2\Gamma^*(T_l))},
\end{equation}
where $J$, the electron transport rate, is equal to $\min{(J_{max}(T_l), J_{\phi}(\phi))}$. The maximum potential electron transport rate, $J_{max}(T_l)$, is given by
\begin{equation} \label{eq:Jmax}
J_{max}(T_l)=J_{max0}\frac{\exp[\frac{H_{aJ}}{RT_0}(1-\frac{T_0}{T_l})]}{1+\exp(\frac{S_{vQ}T_l-H_{dJ}}{RT_l})},
\end{equation}
while the PAR limited electron transport rate, $J_{\phi}(\phi)$, is given by
\begin{equation} \label{eq:radiation}
J_{\phi}(\phi)=\frac{\phi\lambda\kappa_2}{2N_ahc},
\end{equation}
where $\phi$ is the incoming radiation (W/m$^2$), 50 percent of which is considered photosynthetically active radiation (PAR) \citep{Jones1992}, $\lambda$ is the average wavelength (m) for PAR (assumed to be 550 nm), $h$ is Planck's constant (Js), $c$ is the speed of light (m/s), $N_a$ is Avogadro's constant (mol$^{-1}$), and $\kappa_2$ is the quantum yield of photosynthesis in mol CO$_2$ mol$^{-1}$ photons. 

The dark respiration $R_d$ is modelled according to an identical Arrhenius equation with coefficients $R_{d0}$ and $H_{kR}$ (see \citet{Leuning1995, Bartlett2014}).

\subsection{Details of the hydraulic model}\label{sec:AppendixHydro}

\subsubsection*{Water potentials}\label{sec:potentials}
The soil water potential, $\psi_s$, is related to the soil moisture through a strongly nonlinear function given by \citet{Rodriguez2004} and \citet{Daly2004} as
\begin{equation}\label{eq:psis}
\psi_s(s) = \overline{\psi}_s s^{-b},
\end{equation}
where $\overline{\psi}_s$ is the soil water potential at saturation and $b$ is the exponent of the retention curve. The specific plant water capacitance $c$, is defined as the change in relative stored water volume per unit change in water potential ($c=dw/d\psi_w$). In this study we have chosen to approximate the plant water capacitance as constant and the stored water potential $\psi_w$ as a linear function of the relative water storage volume $w$ following \citet{Hunt1991}, i.e.,
\begin{equation}\label{eq:psiw}
\psi_w(w) = \frac{w - 1}{c}.
\end{equation}
This relationship neglects nonlinearities in the pressure-volume relationship caused by osmotic effects at low $w$ in order to include plant water storage with a minimum level of complexity. Although a simplification, this linear relationship is a good approximation in the physically relevant regime for many succulent and CAM species \citep{Nobel1983, Hunt1987, Ogburn2012}. In \textit{O. ficus-indica} specifically, the pressure-volume relationship has been shown to be approximately linear for relative water contents above 20\%; below this point further decreases in relative water content will lead to tissue damage and are not considered reversible \citep{Goldstein1991a}.

\subsubsection*{Conductances}\label{sec:g}

The stomatal conductance to water, $g_{s,H_2O}$, is closely related to the stomatal conductance for CO$_2$ given in Equation \eqref{eq:gms} and is here given by
\begin{equation}\label{eq:gmsH20}
g_{s,H_2O} = 1.6 g_{s,CO_2} + g_{cut},
\end{equation}
where the factor 1.6 accounts for the differences between the the diffusivity in air of CO$_2$ and H$_2$O \citep{Jones1992}. The cuticular conductance, $g_{cut}$, is added to the stomatal conductance to account for the small amount of water vapor lost in the absence of carbon assimilation \citep{Burghardt2003}.

Following \citet{Daly2004}, the soil-root conductance, $g_{sr}$, is assumed to be proportional to the soil hydraulic conductivity, $K(s)$, divided by the average distance from the soil to root surface, i.e.,
\begin{equation}\label{eq:gsr}
g_{sr}(s) = \frac{K(s)\sqrt{RAI_w s^{-d}}}{\pi g\rho_w Z_r},
\end{equation}
where $RAI_w$ is the root area index under well-watered conditions, $s^{-d}$ is a term introduced to model root growth under water-stressed conditions, $g$ is the gravitational constant, and $Z_r$ is the rooting depth. The hydraulic conductivty $K(s)$ is given by
\begin{equation}\label{eq:Ks}
K(s) = K_s s^{2b+ 3},
\end{equation}
where $K_s$ is the saturated hydraulic conductivity and $b$ is a parameter defined in Equation \eqref{eq:psis} (see Table \ref{tab:soilPar}).

The decrease in plant conductance under water stress is modeled by a vulnerability curve so that $g_p$ is close to $g_{pmax}$ for high $\psi_l$ and is close to 0 for low $\psi_l$ due to xylem cavitation \citep{Sperry1998, Daly2004}, i.e.,
\begin{equation}\label{eq:gp}
g_p = g_{pmax}\exp\left[-\left(\frac{-\psi_l}{j}\right)^h\right].
\end{equation}
where $h$ and $j$ are shape parameters.
Following \citet{Waring1978} and \citet{Carlson1991}, the conductance between the water storage tissue and the xylem is assumed to decrease with the fraction of stored water following a power law given by
\begin{equation}\label{eq:gw}
g_w = g_{wmax} w^m,
\end{equation}
where $g_{wmax}$ is the maximum storage-xylem conductance and $a$ is a parameter between 1 and 10, here assumed to be equal to 4. Due to the linear relationship between $w$ and $\psi_w$ imposed by Equation \eqref{eq:psiw}, this assumption is equivalent to assuming a power law relationship between the stored water potential $\psi_w$ and the conductance $g_p$.

\subsubsection*{Hydrology balance with plant water storage}

In the absence of plant water storage, the hydrology balance may be described through Equations \eqref{eq:transair}, \eqref{eq:transplant}, and \eqref{eq:energybal}, which are solved simultaneously for the leaf water potential, $\psi_l$, the leaf temperature, $T_l$, and the transpiration $E$. When plant water storage is included, the formulation of hydrology balance is slightly altered, and a fourth variable, $\psi_x$, is introduced, which describes the water potential at the storage connection node. Equation \eqref{eq:transplant} is now given by 
\begin{equation}\label{eq:transstor}
E = \frac{g_{p}}{1-f}(\psi_x-\psi_l),
\end{equation}
where $\frac{g_p}{1-f}$ is the hydraulic conductance between the storage connection node and the leaf (see Figure \ref{fig:Cap}). Now the hydraulic balance is described by Equations \eqref{eq:transair}, \eqref{eq:energybal}, \eqref{eq:psi_x}, and \eqref{eq:transstor}. To solve this system of equations, Equation \eqref{eq:transstor} is solved for $\psi_x$ and substituted into Equation \eqref{eq:psi_x}, eliminating the unknown $\psi_x$, i.e.
\begin{equation}\label{eq:eliminate}
E = \frac{g_{srfp}(\psi_s-\psi_l)+LAIg_w(\psi_w-\psi_l)}{1 + \frac{g_{srfp}(1-f)}{LAI g_p} + \frac{g_w(1-f)}{g_p}}.
\end{equation} 
The resulting system of three equations - Equation \eqref{eq:transair}, \eqref{eq:energybal}, and \eqref{eq:eliminate} - is now solved simultaneously for the three unknowns,  $\psi_l$, $T_l$, and $E$.

\subsection{Details of the CAM photosynthetic model}\label{sec:AppendixCAM}

The CAM photosynthetic fluxes $A_{sc}$, $A_{sv}$, $A_{vc}$ are modified from \citet{Bartlett2014} to improve model robustness to a range of environmental conditions. The flux $A_{sc}$ from the stomata to the Calvin cycle is given by
\begin{equation}\label{eq:Asc}
A_{sc}(\phi, c_m, T_l, \phi_l, z, M) = \max[(A_{\phi, c_m, Tl}(\phi, c_m, T_l)-R_{dc}(T_l))\times f_{\psi_l}(\psi_l)\times (1-f_C(z, M)),0],
\end{equation}
where $R_{dc}(Tl)$ is the respiration flux to the Calvin cycle and is given by 
\begin{equation}
R_{dc}(T_l) = R_d(T_l)(1-\exp(\phi)).
\end{equation}
The carbon function $f_C(z, M)$ accounts for the circadian rhythm control of the flux dictated by the values of $M$ and $z$ and is given by
\begin{equation}\label{eq:fC}
f_C(z, M) = (1-f_O(z))\frac{M}{\alpha_1M_{max} + M},
\end{equation}
where the order function $f_O(z)$ describes the relative rate of malic acid diffusion from the cell vacuole to the cytoplasm and the overall activation state of the decarboxylation enzymes and is given by
\begin{equation}\label{eq:fO}
f_O(z) = \exp\left[-\left(\frac{z}{\mu}\right)^{c_3}\right],
\end{equation} 
where $\mu$ and $c_3$ are circadian oscillator constants given in Table \ref{tab:CAMPar}. Note that for $f_C = 0$ the flux $A_{sc}$ is the same as the C3 carbon assimilation given in Equation \eqref{eq:Ad}.

The flux $A_{sv}$ from the stomata to the vacuole is adjusted from \citet{Bartlett2014} with a modification which prevents the flux from occurring when light is present and the vacuole is empty. The modified flux is given by
\begin{equation}\label{eq:Asv}
A_{sv}(T_l, \psi_l, z, M) =\left\{
\begin{array}{l l}
0, & \quad \phi > 0$ $ \& $ $M << 1 \\
(A_{sv, max}(T_l) - R_{dv}(T_l))\times f_{\psi_l}(\psi_l) \times f_M(z, M, T_l), & \quad otherwise, \\
\end{array} \right.
\end{equation}
where the malic acid storage function $f_M(z, M, T_l)$ accounts for the circadian control of the flux and is given by
\begin{equation}\label{eq:fM}
f_M(z, M, T_l) = f_O(z) \frac{M_S(T_l)- M}{\alpha_2 M_S(T_l) + (M_S(T_l) - M)},
\end{equation}
where $M_S(T_l)$ is the maximum storage concentration of malic acid and is given by
\begin{equation}
M_S(T_l) = M_{max}\left[\frac{T_H-T_l}{T_H-T_L}(1-\alpha_2)+\alpha_2\right],
\end{equation}
where $T_H$ and $T_L$ are the high and low temperature bounds of the circadian rhythm and $\alpha_2$ is a parameter of the circadian oscillator. $R_{dv}(T_l)$ is the respiration flux to the cell vacuole, given by 
\begin{equation}
R_{dv}(T_l) = R_d(T_l)\exp(-\phi).
\end{equation}
Finally, the flux $A_{vc}$ from the vacuole to the Calvin cycle is given by
\begin{equation}\label{eq:Avc}
A_{vc}(\phi, c_c, T_l, z, M) = (A_{\phi, c_c, T_l} - R_{dc}(T_l))\times f_C(z, M).
\end{equation}


\clearpage

\bibliographystyle{spbasicsimple}


\end{document}